\documentclass[10pt,oneside,onecolumn]{elsart}

\usepackage[ansinew]{inputenc}
\usepackage[english]{babel}
\usepackage{ifthen,shortvrb}
\usepackage[fleqn]{amsmath}
\usepackage{amsxtra,amsfonts,latexsym,amssymb,euscript}
\usepackage{amstext}
\usepackage{graphicx}
\usepackage{dcolumn}
\usepackage{subfigure}
\begin{document}

\begin{frontmatter}
%\journal{Composite Structures, Vol. 95, January 2013, 630–638.}
\title{A multi-physics and multi-scale numerical approach to microcracking and power-loss in photovoltaic modules}
\author{Marco Paggi\corauthref{mp}},
\corauth[mp]{Corresponding author. Tel. +39-011-090-4910 Fax
+39-011-090-4899} \ead{marco.paggi@polito.it}
\author{Mauro Corrado},
\author{Maria Alejandra Rodriguez}
\address{Politecnico di Torino, Department of Structural, Geotechnical and Building
Engineering\\ Corso Duca degli Abruzzi 24, 10129 Torino, Italy}

\begin{abstract}
A multi-physics and multi-scale computational approach is proposed
in the present work to study the evolution of microcracking in
polycrystalline Silicon (Si) solar cells composing photovoltaic (PV)
modules. Coupling between the elastic and the electric fields is
provided according to an equivalent circuit model for the PV module
where the electrically inactive area is determined from the analysis
of the microcrack pattern. The structural scale of the PV laminate
(the macro-model) is coupled to the scale of the polycrystals (the
micro-model) using a multiscale nonlinear finite element approach
where the macro-scale displacements of the Si cell borders are used
as boundary conditions for the micro-model. Intergranular cracking
in the Si cell is simulated using a nonlinear fracture mechanics
cohesive zone model (CZM). A case-study shows the potentiality of
the method, in particular as regards the analysis of the microcrack
orientation and distribution, as well as of the effect of cracking
on the electric characteristics of the PV module.\\

\vspace{0.5cm}\noindent \emph{Notice}: this is the author's version
of a work that was accepted for publication in Composite Structures.
Changes resulting from the publishing process, such as editing,
structural formatting, and other quality control mechanisms may not
be reflected in this document. A definitive version was published in
Composite Structures, Vol. 95, January 2013, 630--638,
DOI:10.1016/j.compstruct.2012.08.014
\end{abstract}
\end{frontmatter}

\section{Introduction}

\begin{figure}
\centering
\includegraphics[width=0.5\textwidth,angle=0]{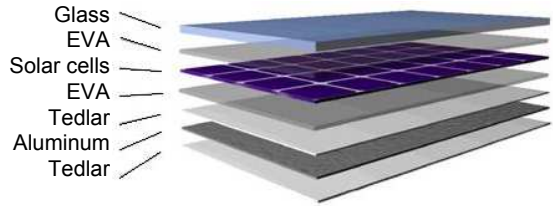}
\caption{Sketch of the composite stack of a PV module.} \label{fig1}
\end{figure}

\begin{figure}
\centering
\includegraphics[width=0.25\textwidth,angle=0]{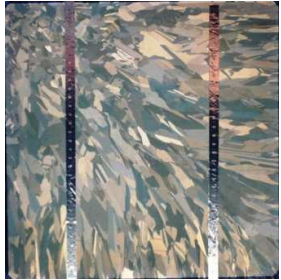}
\caption{Microstructure of a polycrystalline Si cell. The lateral size of the cell is 125 mm.} \label{fig2}
\end{figure}

Photovoltaics (PV) based on Silicon (Si) semiconductors is one the
most growing technology in the World for renewable, sustainable,
non-polluting, widely available clean energy sources. Standard PV
modules are laminates composed of a glass superstrate 4 mm thick, an
encapsulating polymer layer (EVA) 0.5 mm thick, a layer of Si solar
cells 0.166 mm thick, another layer of EVA with the same thickness
as the previous one, and finally a thin multi-layered backsheet made
of Tedlar/Aluminum/Tedlar 0.1 mm thick, see Fig.\ref{fig1}. For more
details about the geometrical and mechanical properties of these
constituent materials, the reader is referred to \cite{PKE11}. The
majority of solar cells available on the market are made of either
mono or polycrystalline Si and are separated by a thin amount of EVA
in their plane. Two main semiconductors, called busbars, connect the
cells together and are placed on the upper and the lower sides of
the cells. The microstructure of a polycrystalline Si cell is shown
in Fig.\ref{fig2}, where we note a significant elongation of the
grains due to production issues.

So far, theoretical and applied research has focused on the increase
of the solar energy conversion efficiency of the cells. Although efficiencies up to $40\%$ have been
reached in the laboratory using single junction GaAs and
multijunction concentrators, the technology based on mono and
multicrystalline Si is still the most competitive on the
market due to the low price of Si semiconductor and the widely
established material processing developed in the field of
electronics \cite{GHS}.

\begin{figure}
\centering
\includegraphics[width=0.5\textwidth,angle=0]{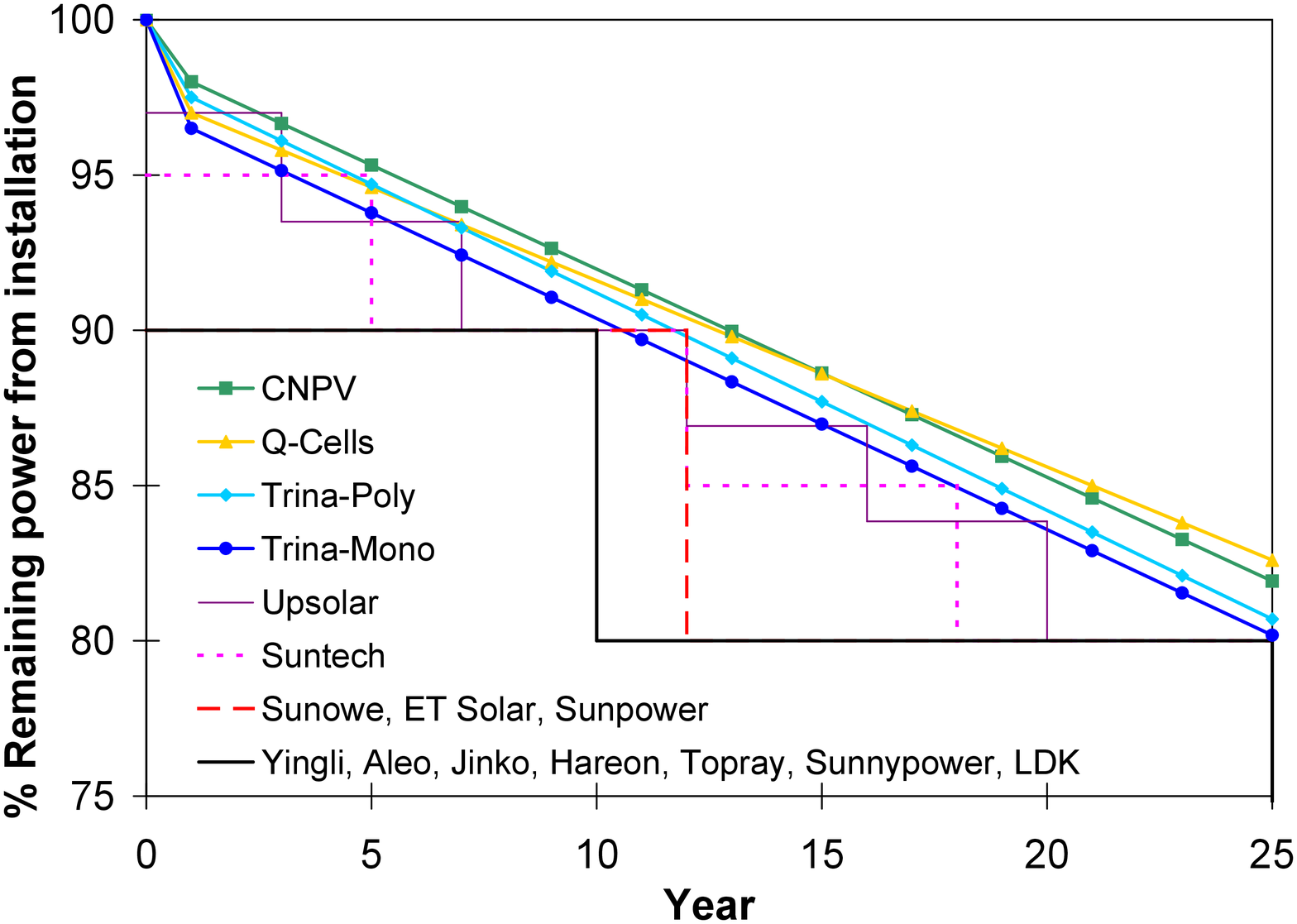}
\caption{Comparison among warranty specifics declared by various producers of PV modules.} \label{fig01}
\end{figure}

Another important issue is the problem of durability, which, however, has received much less attention by the
scientific community so far. In the 1990s, warranties of PV producers allow one to replace PV modules in
case of power losses larger than $10\%$ in the first 10 years, and
then larger than $20\%$ in the next 15 years. The maximum life of PV modules is
considered to be of 25 years. More recently, with more and more
field data of installed modules available, a
linear decreasing performance of the PV module is expected (see the
comparison among various warranty specifics in Fig.\ref{fig01}).

\begin{figure}
\centering
\includegraphics[width=0.5\textwidth,angle=0]{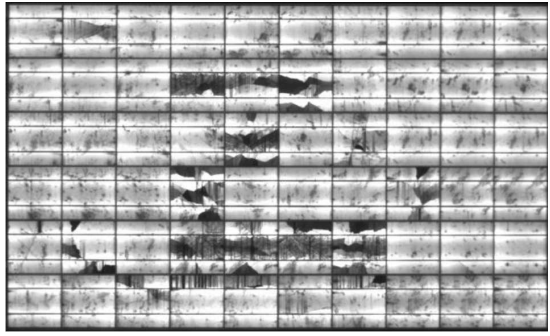}
\caption{Electroluminescence image of a micro-cracked PV module.
Dark regions are electrically inactive areas [4].} \label{fig3}
\end{figure}

The quality control of these composites is of primary concern from
the industrial point of view. On the one hand, the aim is to develop
new manufacturing processes able to reduce the number of cells or
modules rejected by quality control \cite{RBS}. On the other hand,
even if all the damaged cells are theoretically discarded during
manufacturing, it is impossible to avoid the occurrence of
microcracking. Sources of damage in Si cells are transport,
installation and use (in particular impacts, snow loads and
environmental aging caused by temperature and relative humidity
variations). The existing qualification standards IEC 61215 require passing of
severe laboratory tests in a climate chamber. However, microcracking is not used as a quantitative indicator for
the quality assessment of PV modules. Recently, Kajari-Schr\"{o}der et al. \cite{KKEK} have analyzed microcracking resulting from snow
tests and artificial aging in the laboratory using the
electroluminescence technique, see Fig.\ref{fig3}.
Microcracking can lead to large electrically disconnected cell
areas, with up to $16\%$ of power-loss \cite{KKKBB}. In addition to laboratory
tests on single panels, field data published in \cite{WSZB} have shown
that microcracked cells have a non constant current-voltage
characteristics in time and an undesirable increase of the operating temperature.

The aim of the present work is to understand the phenomena leading
to microcracking in Si cells and to quantify the connection between
cracking and power-loss. In this study, an innovative multi-scale
(multi-resolution) and multi-physics numerical method is proposed.
Since Si cells operate in the presence of elastic, thermal and
electric fields, a multi-field (multi-physics) perspective is
considered to be essential to achieve a predictive stage of any
computational model. The multi-resolution approach, on the other
hand, is adopted to simplify the actual 3D problem in a simpler 2D
one, where microcracking in Si cells is numerically simulated under
plane stress conditions and nonlinear fracture mechanics
formulations. To the knowledge of the present authors, the present
model is the first proposed in the literature for the simulation of
microcracking and the resulting prediction of power-loss of PV
modules.

\section{A multi-physics approach}

The study of durability of PV modules requires the characterization
of the effect of microcracking induced by mechanical loads and
thermal excursions on the electric response of the solar panel.
Mechanical loads are induced by vibrations and impacts during
transportation, installation and use of the modules. Deformations
are also induced by the night and day alternating temperature
variation. Moreover, temperature affects
the electric performance of the PV module, since the semiconductor
differential equations present temperature dependent coefficients.
Hence, to achieve a predictive stage, a computational method should
account for the coupling between the elastic, the thermal and the
electric fields. In other words, a multi-field (multi-physics)
approach has to be pursued.

\begin{figure}
\centering
\includegraphics[width=0.7\textwidth,angle=0]{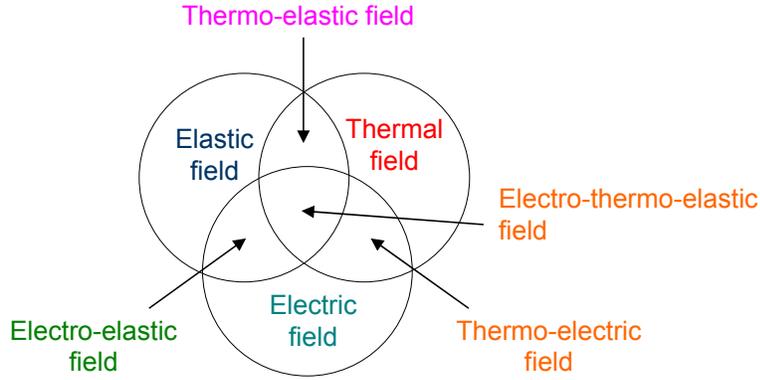}
\caption{Interplay among the elastic, the thermal and the electric
fields in PV modules.} \label{fig4}
\end{figure}

A conceptual diagram of the interplay among the various fields is
shown in Fig.\ref{fig4}. The coupling between the elastic and the
thermal fields can be accounted for by the classic equations of
thermoelasticity. However, in presence of cracking, a specific
treatment of the partial differential equations describing heat
conduction should be considered, as well as its inherent nonlinearities.

\begin{figure}
\centering
\includegraphics[width=0.5\textwidth,angle=0]{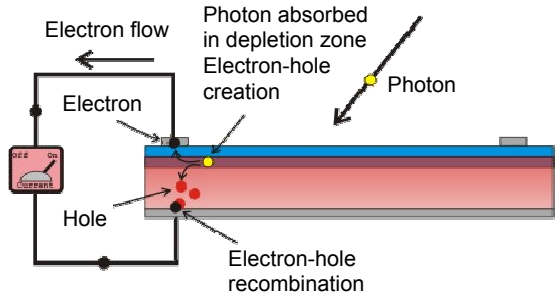}
\caption{Physics of a solar cell.} \label{fig5}
\end{figure}

\begin{figure}
\centering
\includegraphics[width=0.5\textwidth,angle=0]{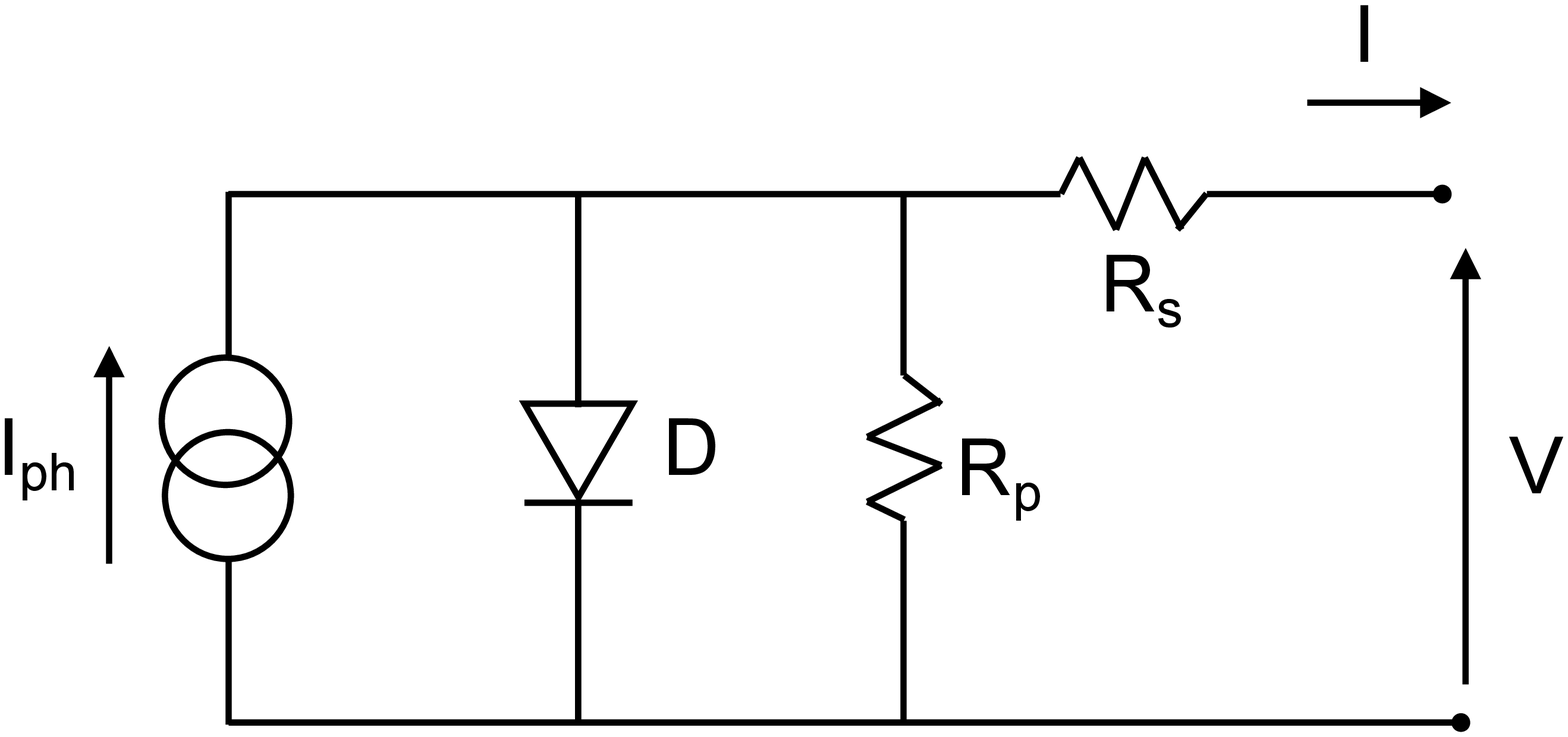}
\caption{Single diode equivalent circuit model.} \label{fig6}
\end{figure}

Regarding the interplay between the electric and the thermal fields,
a rigorous approach should consider the partial differential
equations describing the electric and magnetic fields inside the
semiconductor, according to the physics of the photovoltaic effect
(see Fig.\ref{fig5}).
In the present work, a simplified approach is proposed by
considering an equivalent electric circuit of the cells and of the
module, see Fig.\ref{fig6}. Although mathematically less rigorous
since the interplay of the thermal and elastic fields is accounted
for in a global way, this approach is considered to be particularly
appealing from the industrial point of view, since it can be
implemented in Matlab/Simulink or in commercial finite element (FE)
software as a user-defined subroutine.

The basic idea is to consider the photovoltaic effect as the source
of a photonic current $I_{\text{ph}}$ due to the movement of
electrons and holes in Si as a result of the so-called $p-n$
junction. The $p-n$ junction effect can then be modelled with one or two diodes \cite{GM99}. In the present study, a one single diode model is used.
Series and parallel resistances complete the circuit.
Particular important is the series resistance, which takes into
account the bulk resistance of the semiconductor, contacts and
interconnectors. The parallel resistance, also called \emph{shunt
resistance}, is used to model the effect of impurities and non
idealities of the $p-n$ junction.

The equation describing the current-voltage response of a PV module
with Si cells connected in series is \cite{GM99}:
\begin{equation}\label{eqcir_1}
I=I_{\text{ph}}-I_{\text{s}}\left\{\exp{\left[\dfrac{e(V/n+IR_{\text{s}})}{a\kappa T}\right]}-1\right\}-\dfrac{V/n+IR_{\text{s}}}{R_{\text{p}}},
\end{equation}
where $n$ is the number of cells, $I_{\text{ph}}$ (A) is the photonic current, $I_{\text{s}}$ (A) is the saturation current, $V$ (V) is the module terminal voltage, $I$ (A) is the module terminal current, $R_{\text{s}}$ ($\Omega$) is the series resistance, $R_{\text{p}}$ ($\Omega$) is the parallel resistance, $e=1.6\times 10^{-19}$ C is the electronic charge, $a\cong 2$ is the diode quality factor for polycrystalline Si, $\kappa=1.38\times 10^{-23}$ J/K is the Boltzmann's constant, and $T$ (K) is the ambient temperature.

The main quantities in Eq.\eqref{eqcir_1} depend on the cell temperature $T$ \cite{GM99}:
\begin{subequations}
\begin{align}
I_{\text{ph}}&=I_{\text{ph}}^{T=300}\left[1+k_0(T-300)\right],\\
I_{\text{s}}&=k_1 T^3 \exp{\left(-\dfrac{e V_g}{\kappa T}\right)},\\
R_{\text{s}}&=R_{\text{s}}^{T=300} \left[1-k_2(T-300)\right],\\
R_{\text{p}}&=R_{\text{p}}^{T=300}\exp{\left(-k_3 T\right)},
\end{align}
\end{subequations}
where $V_g$ (V) is the band gap voltage and the coefficients $k_i$ are determined from experiments \cite{GM99}.

From the mathematical point of view, Eq.\eqref{eqcir_1} is implicit
and nonlinear. Hence, for a given value of voltage $V$, its solution
has to be obtained using a numerical method. Here, the
Newton-Raphson method is adopted to achieve a quadratic convergence of the numerical scheme. After a suitable
manipulation, Eq.\eqref{eqcir_1} becomes:
\begin{equation}\label{eqcir_2}
f(I)=I-I_{\text{ph}}+I_{\text{s}}\left\{\exp\left[{\dfrac{e(V/n+IR_{\text{s}})}{a\kappa T}}\right]-1\right\}+\dfrac{V/n+IR_{\text{s}}}{R_{\text{p}}}=0.
\end{equation}

At the generic iteration $i+1$, the solution is sought from the
approximation at the previous step $i$:
\begin{equation}\label{eqcir_3}
I_{i+1}=I_{i}-\left[\dfrac{\text{d}f}{\text{d}I}\right]^{-1}_{I_i}\,
f(I_i),
\end{equation}
where the derivative $\text{d}f/\text{d}I$ is:
\begin{equation}\label{eqcir_4}
\dfrac{\text{d}f}{\text{d}I}=1+I_{\text{s}}\exp{\left[\dfrac{e(V/n+IR_{\text{s}})}{a\kappa T}\right]}\dfrac{e R_{\text{s}}}{a\kappa T}+\dfrac{R_{\text{s}}}{R_{\text{p}}}.
\end{equation}
Equation \eqref{eqcir_3} is iterated until convergence, i.e., until
$|I_{i+1}-I_{i}|<\text{tol}$, where $\text{tol}=0.001$ is a used
prescribed tolerance. The state $I_0=I_{\text{SC}}$ is selected as
the starting point for the iteration procedure.

\begin{figure}
\centering
\includegraphics[width=0.6\textwidth,angle=0]{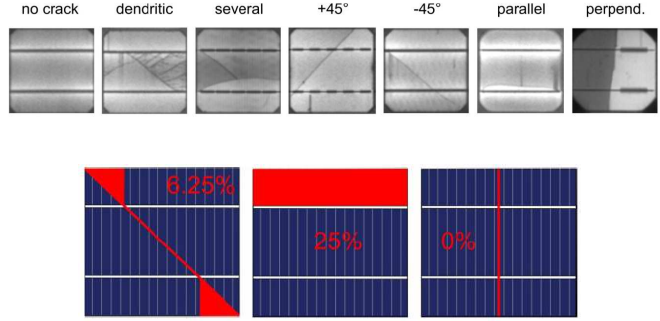}
\caption{Representation of electrically inactive cell areas, adapted
from \cite{KKKBB}.} \label{fig7}
\end{figure}

To model the effect of cracking on the electric
field, we note that the the saturation current $I_{\text{s}}$ is linearly dependent on the electrically active cell area, as found in \cite{AR}. Electrically inactive cell areas are those whose electric flux directed towards the two main conductors connecting the cells (busbars) is interrupted by a crack. Examples are shown in Fig.\ref{fig7}, where we note that cracks parallel to the busbar are the most dangerous ones, whereas cracks perpendicular to the busbar might have a negligible effect on the $I-V$ characteristics \cite{KKKBB}, since they do not interrupt the electron flux. Intermediate situations can occur and are related to intermediate inclinations of microcracks. According to this criterion, we introduce a damage variable for each cell defined as follows:
\begin{equation}
D=\dfrac{A_{\text{inactive}}}{A_{\text{total}}},
\end{equation}
where $A_{\text{inactive}}$ and $A_{\text{total}}$ are,
respectively, the inactive and the total cell areas. The same
definition applies also to an ensemble of cells in series (a PV
module). In this case the damage variable is the maximum of the
damage variables of the various cells. As a result of damage, the
saturation and photonic currents become:
\begin{subequations}
\begin{align}
I_{\text{s}}&=I_{\text{s}}^{D=0}(1-D),\\
I_{\text{ph}}&=I_{\text{ph}}^{D=0}(1-D).
\end{align}
\end{subequations}
In general, the graph of the output current $I$ as a function of voltage $V$ is almost constant and equal to the short circuit current $I_0=I_{\text{sc}}$ for a voltage less than the open circuit voltage $V_{\text{oc}}$. For $V\to V_{\text{oc}}$, the $I-V$ curve is rounded up and the current comes to zero for $V=V_{\text{oc}}$. The output power is equal to the instantaneous product of the PV current intensity and the PV voltage, $P=I\,V$. Due to the shape of the $I-V$ curve, the $P-V$ characteristics is linear up to a maximum, called \emph{maximum power point}, and then comes down to zero for $V=V_{\text{oc}}$. The electric performance can be synthetically described by the \emph{fill factor}:
\begin{equation}
FF=\dfrac{V_{\text{MP}}I_{\text{MP}}}{V_{\text{oc}}I_{\text{sc}}},
\end{equation}
where $V_{\text{MP}}$ and $I_{\text{MP}}$ are the voltage and the current of the maximum power point. Hence, the power-loss due to microcracking can be estimated by comparing the fill factors of the intact module and of the damaged one.

\section{A multi-scale (multi-resolution) computational method}

\begin{figure}
\centering
\includegraphics[width=0.65\textwidth,angle=0]{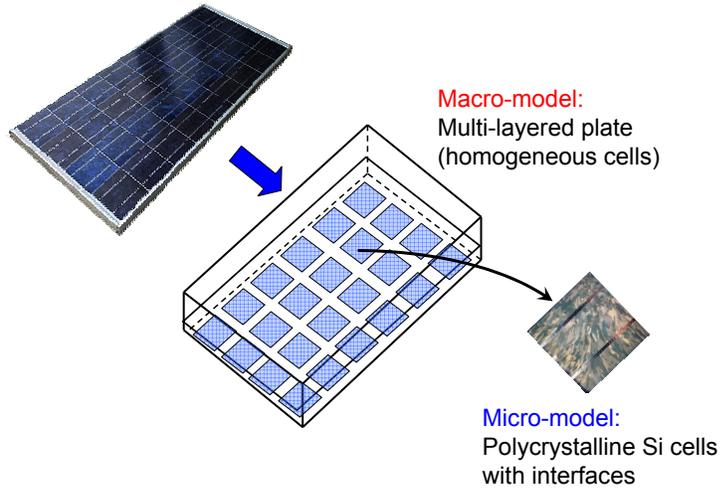}
\caption{A real PV module and its macro- and micro-models.}
\label{fig8}
\end{figure}

Photovoltaic modules are laminated composites where the thickness of
the various layers are very different from each other, as outlined
in the Introduction. Moreover, Si cells are separated by a thin
interspace of EVA and are made of a polycrystalline material whose
microstructure has to be considered to predict microcracking. These
complexities suggest the use of 3D methods, which are however
computationally expensive. In the present study, a simplified
approach to reduce the computational complexity is proposed and it
is based on two levels of resolution, see Fig.\ref{fig8}.

The macro-model consists in a laminated composite analyzed in the
framework of the small-deflection theory with homogeneous Si cells.
The layers are considered to be fully bonded along their interfaces.
The grain boundaries in the material microstructure of Si cells and
the related intergranular fracture are neglected here. The weak form
for the macro-model problem is:
\begin{equation}\label{wf1}
\delta W=-\int_{\Omega} \delta
\left(\mbox{\boldmath{$\nabla$}}w\right)^{\text{T}}\mbox{\boldmath{$D$}}\mbox{\boldmath{$\nabla$}}w\,\text{d}\Omega+\int_{\Omega}
q\delta w\,\text{d}\Omega,
\end{equation}
with simple support boundary conditions on the border of the plate:
\begin{equation}
w(\partial \Omega)=0.
\end{equation}
where $\Omega$ and $\partial\Omega$ are, respectively, the laminate
plate and its border. The variable $w$ is the transversal
displacement that, together with the vector
$\mbox{\boldmath{$\theta$}}=[\theta_x,\,\theta_y]^{\text{T}}$
containing the nodal rotations in directions $X$ and $Y$, will
represent the unknown generalized displacements of the continuum. To
simulate the snow test prescribed by the standard IEC 61215 for the
qualification of PV modules, a uniform transversal pressure $q$ is
imposed to the upper surface of the PV module.

The operator $\mbox{\boldmath{$\nabla$}}$ in Eq.\eqref{wf1} is given by:
\begin{equation}
\mbox{\boldmath{$\nabla$}}=\left(\dfrac{\partial^2}{\partial^2
x},\,\dfrac{\partial^2}{\partial^2 y},\,2\dfrac{\partial^2}{\partial
x
\partial y}\right)^{\text{T}}.
\end{equation}
The constitutive matrix $\mbox{\boldmath{$D$}}$ is
\begin{equation}
\mbox{\boldmath{$D$}}=K\left[\begin{array}{ccc}
          1 & \nu & 0 \\
          \nu & 1 & 0 \\
          0 & 0 & \dfrac{1-\nu}{2}
        \end{array}\right]
\end{equation}
and the coefficient $K$ of the multi-layered plate depends on the layer arrangement \cite{SR74}:
\begin{subequations}\label{rig}
\begin{align}
K&=\dfrac{AC-B^2}{A},\\
A&=\sum_k \dfrac{E_k}{1-\nu_k^2}(z_k-z_{k-1}),\\
B&=\sum_k \dfrac{E_k}{1-\nu_k^2}\dfrac{z^2_k-z^2_{k-1}}{2},\\
C&=\sum_k \dfrac{E_k}{1-\nu_k^2}\dfrac{z^3_k-z^3_{k-1}}{3},
\end{align}
\end{subequations}
where $k$ is the total number of layers, $z_k$ is the absolute value
of the distance of the lower interface of the considered layer from
the upper side of the plate, and $E_k$ and $\nu_k$ are the Young's
modulus and the Poisson's ratio of the $k-$th layer (see Fig. 10).
Since Si cells are not continuous in the plate plane, two different
values of $K$ will be used for the finite elements belonging to the
portions of the plates with Si cells and without Si cells (see also
Fig.\ref{fig1} for a visual representation).

\begin{figure}
\centering
\includegraphics[width=0.8\textwidth,angle=0]{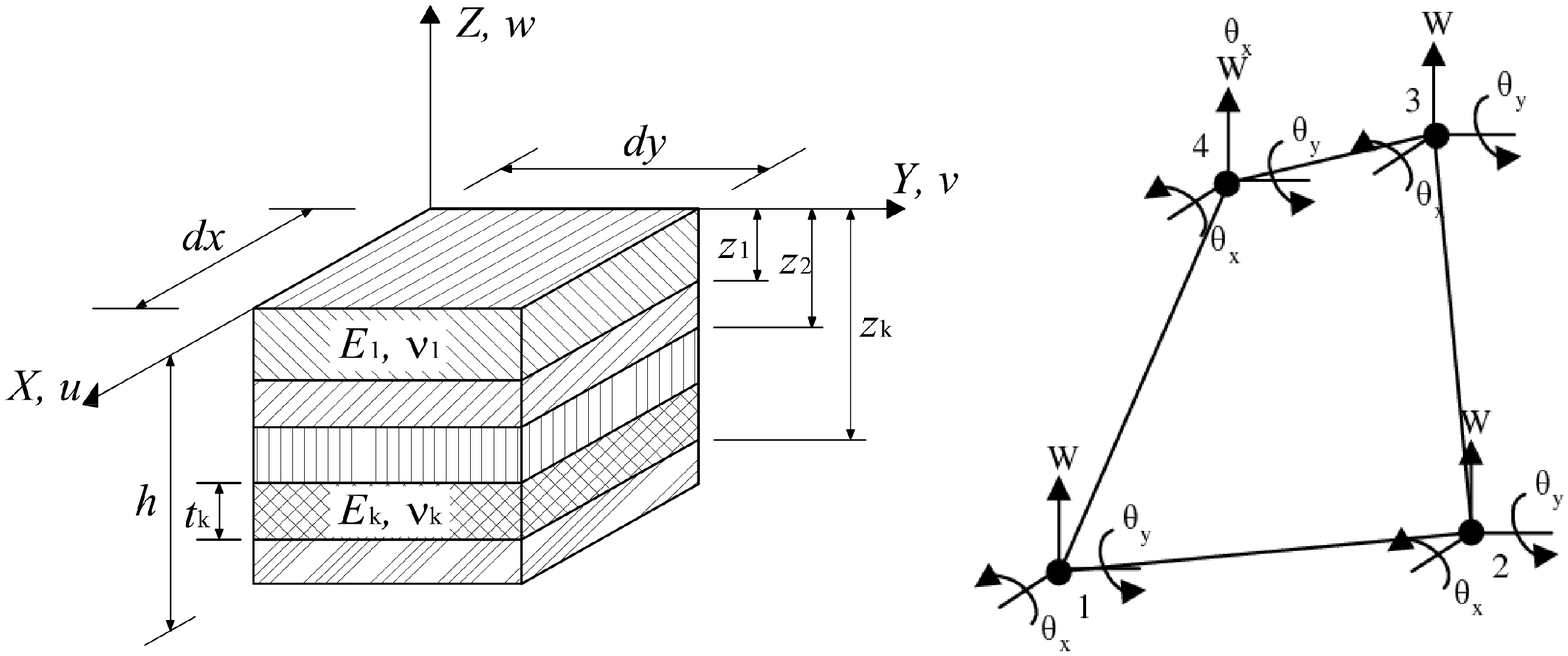}
\caption{The plate element used in the macro-model and its degrees of freedom.} \label{figsc}
\end{figure}

The FE implementation of the macro-model is done by using standard
linear elastic plate elements, see Fig.\ref{figsc}. Using quadrilater finite elements with linear shape
functions, the unknown displacements and rotations are discretized with the use of
standard interpolation functions:
\begin{subequations}
\begin{align}
w&=\mbox{\boldmath{$N$}}\mbox{\boldmath{$\eta$}},\\
\mbox{\boldmath{$\theta$}}&=\mbox{\boldmath{$N$}}_\theta\mbox{\boldmath{$\eta$}},
\end{align}
\end{subequations}
where
$\mbox{\boldmath{$\eta$}}=[w_1,\theta_{x1},\theta_{y1},...,w_4,\theta_{x4},\theta_{y4}]^{\text{T}}$
and:
\begin{subequations}
\begin{align}
\mbox{\boldmath{$N$}}&=[N_1,0,0,...,N_4,0,0],\\
\mbox{\boldmath{$N$}}_\theta &=\left[
                             \begin{array}{ccccccc}
                               0 & N_1 & 0 & ... & 0 & N_4 & 0 \\
                               0 & 0 & N_1 & ... & 0 & 0 & N_4 \\
                             \end{array}
                           \right].
\end{align}
\end{subequations}

Introducing these relations in the weak form \eqref{wf1}, we have:
\begin{equation}
\delta W=\delta
\mbox{\boldmath{$\eta$}}^{\text{T}}\left(-\int_{\Omega}
\left(\mbox{\boldmath{$\nabla$}}\mbox{\boldmath{$N$}}\right)^{\text{T}}\mbox{\boldmath{$D$}}\mbox{\boldmath{$\nabla$}}\mbox{\boldmath{$N$}}\,\text{d}\Omega\,\mbox{\boldmath{$\eta$}}+\int_{\Omega}
\mbox{\boldmath{$N$}}^{\text{T}}q\,\text{d}\Omega\right),
\end{equation}
and setting it equal to zero, for the arbitrariness of $\delta
\mbox{\boldmath{$\eta$}}$, the standard expressions for the stiffness matrix, $\mbox{\boldmath{$K$}}$, and for the external load vector, $\mbox{\boldmath{$F$}}$,  are obtained:
\begin{equation}
\left(\int_{\Omega}
\left(\mbox{\boldmath{$\nabla$}}\mbox{\boldmath{$N$}}\right)^{\text{T}}\mbox{\boldmath{$D_\text{A}$}}\mbox{\boldmath{$\nabla$}}\mbox{\boldmath{$N$}}\,\text{d}\Omega\right)\,\mbox{\boldmath{$\eta$}}=\int_{\Omega}
\mbox{\boldmath{$N$}}^{\text{T}}q\,\text{d}\Omega \quad \Rightarrow \quad \mbox{\boldmath{$K$}}\,\mbox{\boldmath{$\eta$}}=\mbox{\boldmath{$F$}}.
\end{equation}

The displacements $u$ (in the $x$ direction) and $v$ (in the $y$ direction) in the plane of the cells are computed from the nodal rotations as follows:
\begin{subequations}
\begin{align}
u &=\theta_y\,z,\\
v &=-\theta_x\, z,
\end{align}
\end{subequations}
where $z=-1.953$ mm is the distance from the center of mass of the cross-section to the Si cell plane.

It is notable to remark that these displacements impose a tensile
stress state to the cells, since they are positioned far below the
neutral axis of a generic 2D cross-section (see \cite{PKE11} for an
analysis on the overall stiffness contribution of the various
layers).

In the multi-scale approach, the in-plane displacements at the boundary of each cell are
transferred to the micro-model, where a higher resolution of analysis is
considered. Namely, the material microstructure of polycrystalline Si
is taken into account and the progress of intergranular decohesion
at grain boundaries is analyzed under plane stress conditions. Node matching at the cell boundaries for the macro-scale and micro-scale FE meshes is not necessary, since linear interpolation is used to project the boundary displacements from the regularly spaced macro-scale FE mesh to the non regular micro-scale mesh.

In the present modelling, it is implicitly assumed that the in plane displacements are responsible
for cracking and that bending configuration of the cell can be
neglected, so that the deformed configuration is confused with the
undeformed one in the micro-model. As a result of these approximations, the principle of virtual work for the micro-model reads:
\begin{equation}\label{weak_int}
\int_{V}\left(\mbox{\boldmath{$\nabla$}}\delta \mbox{\boldmath{$u$}}\right)^{\text{T}}\mbox{\boldmath{$\sigma$}}\,\text{d}V-
\int_{S}\delta\mbox{\boldmath{$g$}}^{\text{T}}\mbox{\boldmath{$t$}}\,\text{d}S=
\int_{\partial V}\delta\mbox{\boldmath{$u$}}^{\text{T}}\mbox{\boldmath{$f$}}\,\text{d}S,
\end{equation}
where the first term on the l.h.s. is the classical virtual work of deformation of the bulk $V$ and the r.h.s. is the virtual work of the tractions acting on the boundaries of the cell $\partial V$. The second term on the l.h.s. is the contribution to the virtual work of the interface normal and tangential cohesive tractions $\mbox{\boldmath{$t$}}=(\tau,\sigma)^{\text{T}}$ for the corresponding relative sliding and opening displacements $\mbox{\boldmath{$g$}}=(g_{\text{T}},g_{\text{N}})^{\text{T}}$ at grain boundaries $S$. According
to the cohesive zone model (CZM), tractions normal and tangential to the interface are
opposing to the relative opening and sliding displacements of the
grains evaluated at the interface level. In the present study, the
Mixed Mode formulation by Tvergaard \cite{T} is adopted, since it is
suitable for modelling grain boundary decohesion in polycrystalline
materials. The nonlinear equations relating the cohesive tractions
to the normal and tangential relative displacements, $g_\text{N}$
and $g_\text{T}$, are:
\begin{align}
\sigma &=\dfrac{g_\text{N}}{l_{\text{Nc}}}P(\lambda),\\
\tau &=\gamma\dfrac{g_\text{T}}{l_{\text{Tc}}}P(\lambda),
\end{align}
where $P(\lambda)=27\sigma_{\max}(1-2\lambda+\lambda^2)/4$ and $\lambda=\sqrt{\left(g_\text{N}/l_{\text{Nc}}\right)^2+\left(g_\text{T}/l_{\text{Tc}}\right)^2}$.
The FE discretization of the micro-model is performed by using linear triangular
elements for the grains and linear interface elements for the
interfaces \cite{PW2,PW3}. The discretized interface contribution in Eq.\eqref{weak_int} becomes:
\begin{equation}\label{weak_int2}
\Delta \delta W_{\text{int}}=
\delta\mbox{\boldmath{$u$}}^{\text{T}}\mbox{\boldmath{$R$}}^{\text{T}}
\int_{S}\mbox{\boldmath{$B$}}^{\text{T}}\mbox{\boldmath{$t$}}\,\text{d}S,
\end{equation}
where $\delta \mbox{\boldmath{$u$}}=[u_1,v_1,\dots,u_4,v_4]^{\text{T}}$ is the displacement vector, $\mbox{\boldmath{$R$}}$ is the rotation matrix of the interface element, and the matrix $\mbox{\boldmath{$B$}}$ contains its shape functions:
\begin{equation}
\mbox{\boldmath{$B$}}=\left[
                        \begin{array}{cccccccc}
                          -N_1 & 0 & -N_2 & 0 & N_2 & 0 & N_1 & 0 \\
                          0 & -N_1 & 0 & -N_2 & 0 & N_2 & 0 & N_1 \\
                        \end{array}
                      \right]
\end{equation}

Due to the nonlinearity of the interface constitutive relation given
by the CZM, the Newton-Raphson scheme is used, which allows to achieve a quadratic
convergence in the computation. The reader is referred to \cite{PW2} for more details about the computational issues. In this context, linearization of Eq.\eqref{weak_int2} yields:
\begin{equation}\label{weak_int3}
\Delta \delta W_{\text{int}}=
\delta\mbox{\boldmath{$u$}}^{\text{T}}\mbox{\boldmath{$R$}}^{\text{T}}
\int_{S}\mbox{\boldmath{$B$}}^{\text{T}}\mbox{\boldmath{$C$}}\mbox{\boldmath{$B$}}\mbox{\boldmath{$R$}}\mbox{\boldmath{$u$}}\,\text{d}S,
\end{equation}
where $\mbox{\boldmath{$C$}}$ is the tangent constitutive matrix of the interface element containing the partial derivatives of the cohesive tractions w.r.t. the opening and sliding relative displacements \cite{PW2}.

\begin{figure}
\centering
\includegraphics[width=0.7\textwidth,angle=0]{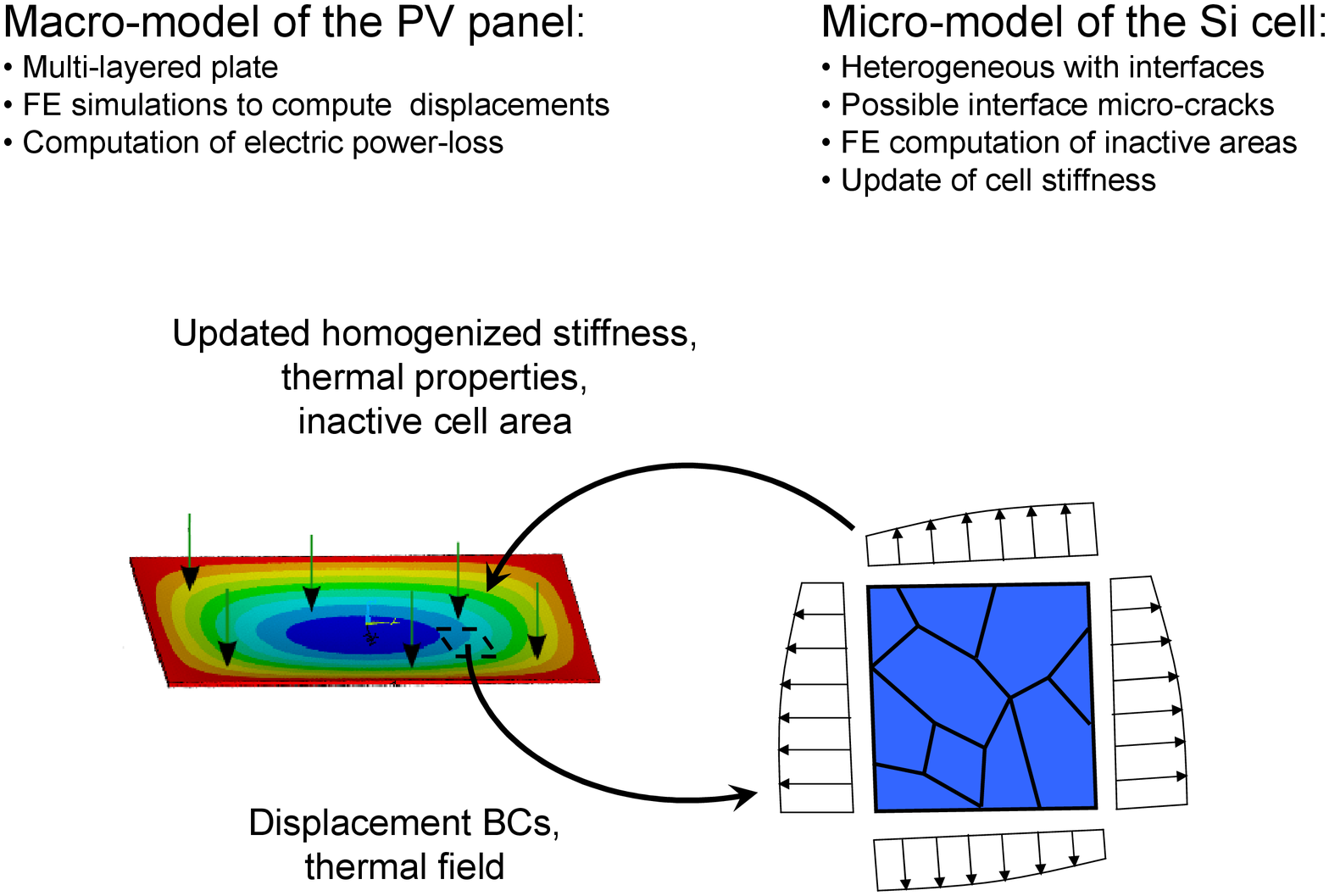}
\caption{Iterative procedure involving the macro- and micro-models.}
\label{fig9}
\end{figure}

In the micro-model, microcracking is originated by the in plane
displacements passed as input from the macro-model (see
Fig.\ref{fig9}). This allows us to compute the updated stiffness of
the Si cell, which coincides with that used in the macro-model at the first iteration
 only in absence of cracking. Moreover, the
electrically inactive cell areas are determined from the inspection
of the crack pattern. This updated information is passed back to the
macro-model where the problem is solved again with the updated
constitutive matrices. This procedure is iterated until convergence
in the computed macro-displacement field is achieved. Since cracking
affects only the Si cells, whose contribution to the overall
stiffness of the plate element is small as compared to the the other
layers, convergence is very fast.

This solution scheme implies the uncoupling between the
microstructures of the Si cells analyzed in the micro-scale computations. This additional
source of error, which is however limited by the fact that Si cells
are not continuous but separated by EVA interlayers, is also
minimized in the iterative convergence scheme. Regarding the
computation of the tangent Young's modulus of the homogenized Si
polycrystalline to be used in the macro-model computations,
different techniques could be invoked. In the present study, the Young's modulus of Si is estimated by the ratio between the average stress and strain in the micro-model of the Si cell. The Poisson's ratio at the macro-level is not updated, since we do not expect significant variations due to microcracking.

\section{A numerical example}

In this section, a numerical example showing the applicability of the proposed numerical method to a realistic case study is proposed. The aim is the determination of microcracking in Si cells, the quantification of the electrically inactive cell areas and finally the computation of the $I-V$ and $P-V$ characteristics of the PV module. The comparison between the characteristics of the intact and microcracked modules will provide a measure of power-loss due to cracking.

\begin{figure}
\centering
\includegraphics[width=0.7\textwidth,angle=0]{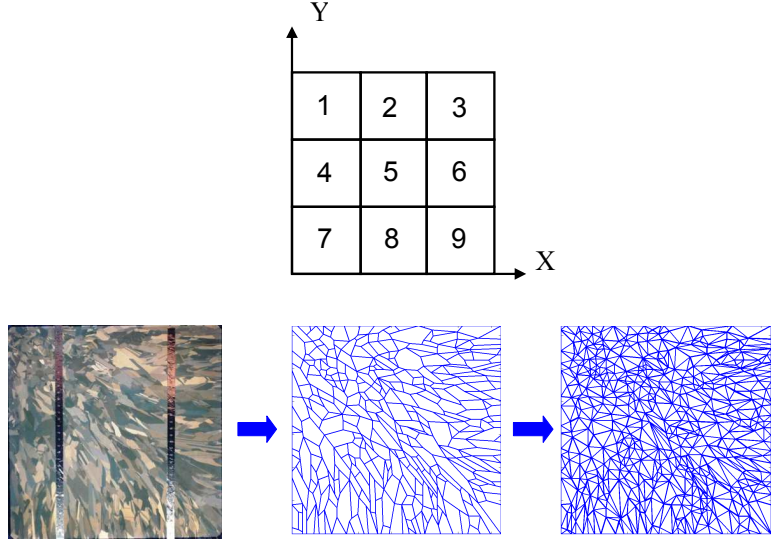}
\caption{A PV module with $3\times 3$ cells. Sequence of
operations for the generation of the FE mesh of a Si cell for the
micro-model: starting from a photo of the microstructure, the grain boundaries
are identified and the FE mesh with interface elements is generated.} \label{fig10}
\end{figure}

A square PV module composed of $3\times 3$ polycrystalline Si cells
subjected to a uniform pressure $q=5400$ N and simply supported
along its sides is analyzed (see Fig.\ref{fig10} for the numeration
of the Si cells). This type of load is recommended in the
qualification standards to simulate the effect of a heavy snow. All
the geometrical and mechanical parameters used for the plate
elements in correspondence of Si cells and their separating domains
are reported in Tab. 1 and 2, respectively. The flexural rigidity
coefficient $K$ of the FE inside the cells is equal to 552746 Nmm,
whereas in the separating domains is equal to 412138 Nmm. It is
worth noting that the approximation of the PV panel as a homogeneous
glass plate by disregarding the stiffening contributions of the
layers below it, as considered in \cite{KKEK}, leads to a flexural
rigidity coefficient $K=409,135$ Nmm. Such a value is considerably
lower than the flexural rigidities used by accounting all the
layers, with a consequent over-estimation of the displacements.

\begin{table}\label{tab1}
\begin{center}
\begin{tabular}{|l|r|r|r|r|}\hline
  Layer & $H_k$    & $z_k$  & $E_k$   & $\nu_k$ \\
        &   (mm) & (mm) & (MPa) & ($-$) \\\hline
Glass   &   4.000 & 4.000 & 73000   & 0.22\\
EVA     &   0.500 & 4.500 & 10      & 0.10\\
Si      &   0.166 & 4.666 & 130000  & 0.22\\
EVA     &   0.500 & 5.166 & 10      & 0.10\\
Backsheet & 0.100 & 5.266 & 2800    & 0.10\\\hline
\end{tabular}\caption{Geometrical and mechanical parameters for FE elements inside Si-cells (macro-scale model).}
\end{center}
\end{table}

\begin{table}\label{tab2}
\begin{center}
\begin{tabular}{|l|r|r|r|r|}\hline
  Layer & $H_k$    & $z_k$  & $E_k$   & $\nu_k$ \\
        &   (mm) & (mm) & (MPa) & ($-$) \\\hline
Glass   & 4.000 &   4.000 & 73000 & 0.22\\
EVA & 1.166 &   5.166 & 10  & 0.10\\
Backsheet   & 0.100 &   5.266 & 2800 &  0.10\\\hline
\end{tabular}\caption{Geometrical and mechanical parameters for FE elements inside the domain between two adjacent cells (macro-scale model).}
\end{center}
\end{table}

For the micro-model, the microstructure of the Si cells is obtained from the real image in Fig.\ref{fig2}. Although a statistical variability of the grain size distribution and orientation takes place from a cell to another, we prefer here to use the same microstructure for all the cells. In absence of fracture anisotropy, the boundary cell displacements computed from the macro-model should induce the same crack pattern in the cells n. 1, 3, 7 and 9. The same reasoning applies to the cells n. 2, 4, 6 and 8. Hence, the comparison between different crack patterns in the various cells will put into evidence the role of the orientation of the grain boundaries.

The generation of the FE meshes for the micro-model is not a trivial
task and a specific pre-processor developed in house is used. First,
starting from a photo of the material microstructure, the grain
boundaries are identified. Afterwards, all the grains are shrunk
with respect to their center of mass and the interface nodes
defining their polygonal geometries are duplicated. At this point,
all the grains are meshed in their interior and interface elements
are placed along the grain boundaries. The resulting FE connectivity
matrices are given as input to the finite element analysis programme
FEAP \cite{feap}, where cohesive interface elements have been
implemented by the first author \cite{PW2}. The properties of the
CZM are chosen to represent the fracture properties of an
EVA-incapsulated Si ($E=169$ GPa, $\nu=0.22$, $\sigma_{\max}=190$
MPa, $\gamma=1$, $l_{\text{Nc}}=l_{\text{Tc}}=0.0156$ mm).

\begin{figure}
\centering
\includegraphics[width=0.6\textwidth,angle=0]{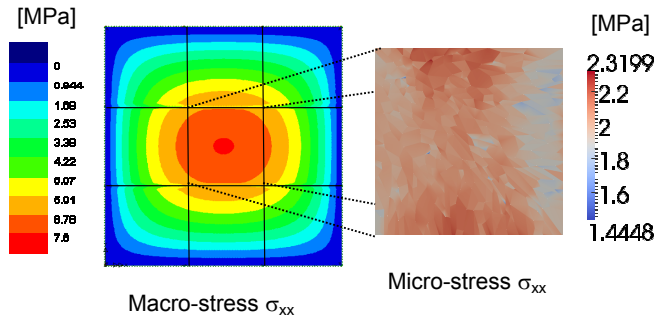}
\caption{Macro- and micro-stresses in the horizontal direction.}
\label{fig11}
\end{figure}

After applying the uniform pressure $q$ to the simply supported PV
module, the macro-micro iterative scheme is applied and, at
convergence, the macro and micro stress fields are computed. The
contour plot of the macro-stress $\sigma_{xx}$ (the stresses in the
horizontal direction $x$) in the PV module and the corresponding
maximum micro-stress field in the central cell is shown in
Fig.\ref{fig11}. As expected, due to microcracking, the micro-stress
field components are in general lower than the macro-stress ones,
for the same applied cell boundary displacements.

\begin{figure}
\centering
\includegraphics[width=0.5\textwidth,angle=0]{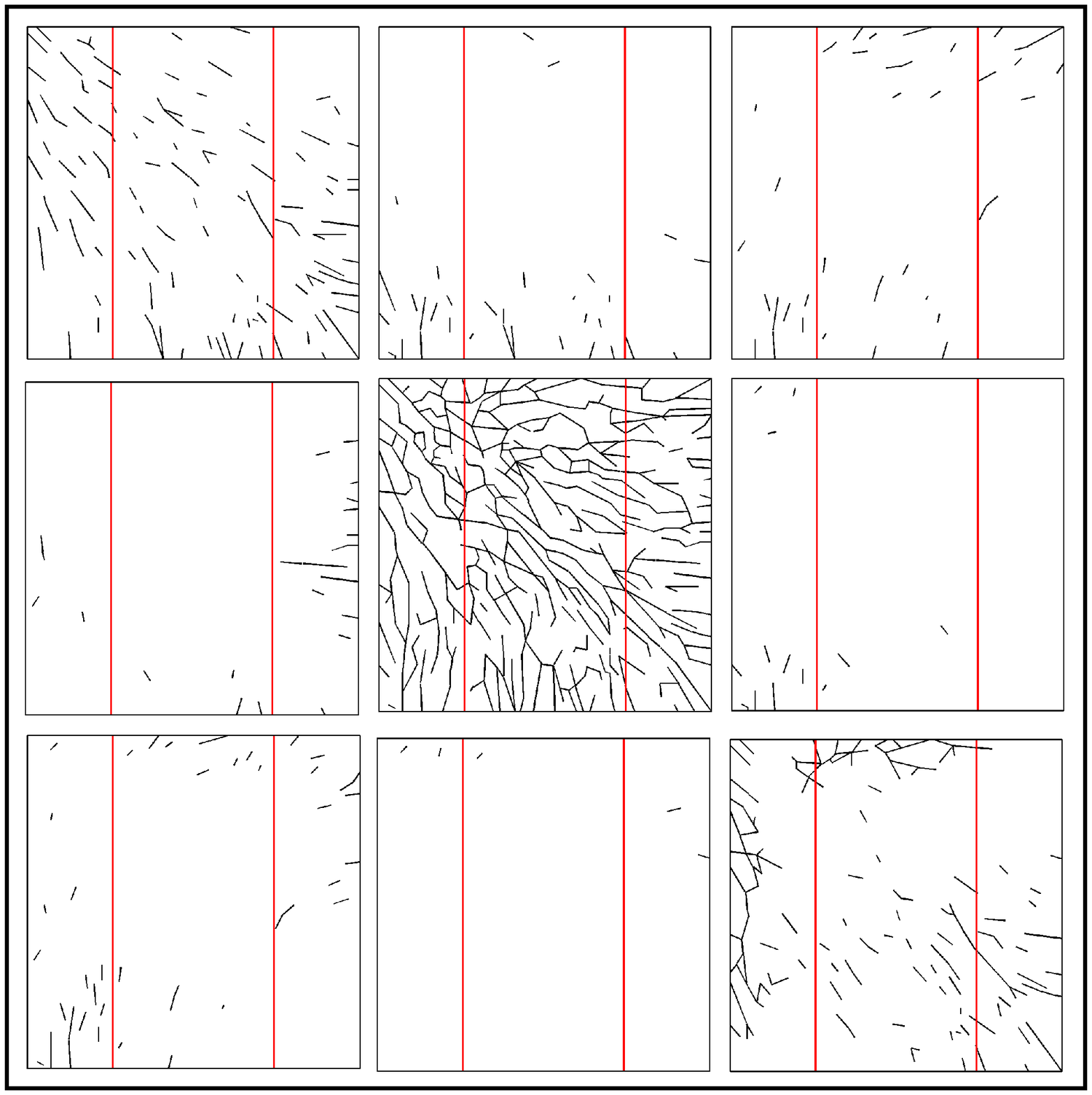}
\caption{Computed crack pattern for a uniform pressure acting on the simply-supported PV module.} \label{fig12}
\end{figure}

The obtained microcrack pattern is shown in Fig.\ref{fig12}. All the lines correspond to interface cracks with $\lambda>2.8\times 10^{-3}$, which is a sufficient low value to select all the microcracks present in the system. However, none of them has $\lambda>1$, i.e., no stress-free macrocracks are present. This is consistent with the experimental evidence showing that, although no cracks can be observed with naked eyes, their effect on the electric performance of the module is quite relevant, as shown in Fig.\ref{fig3} by the electrically inactive cell areas detected using the electroluminescence technique.

Particularly interesting is the analysis of the distribution of the
orientation of the microcracks with respect to the busbars (shown
with red lines in Fig.\ref{fig12}). To do so, the absolute angle of
inclination of the numerically detected microcracks, $|\vartheta|$,
is computed and the corresponding frequency of occurrence is
determined. The angle $|\vartheta|=0^\circ$ denotes cracks
perpendicular to the busbars, whereas cracks with
$|\vartheta|=90^\circ$ are parallel to the busbars. The latter lead
to the highest electrically inactive cell areas and are therefore
particularly harmful.

\begin{figure}
\centering \mbox{\subfigure[Central vs. down right corner cells]{\includegraphics[width=.4\textwidth,angle=0]{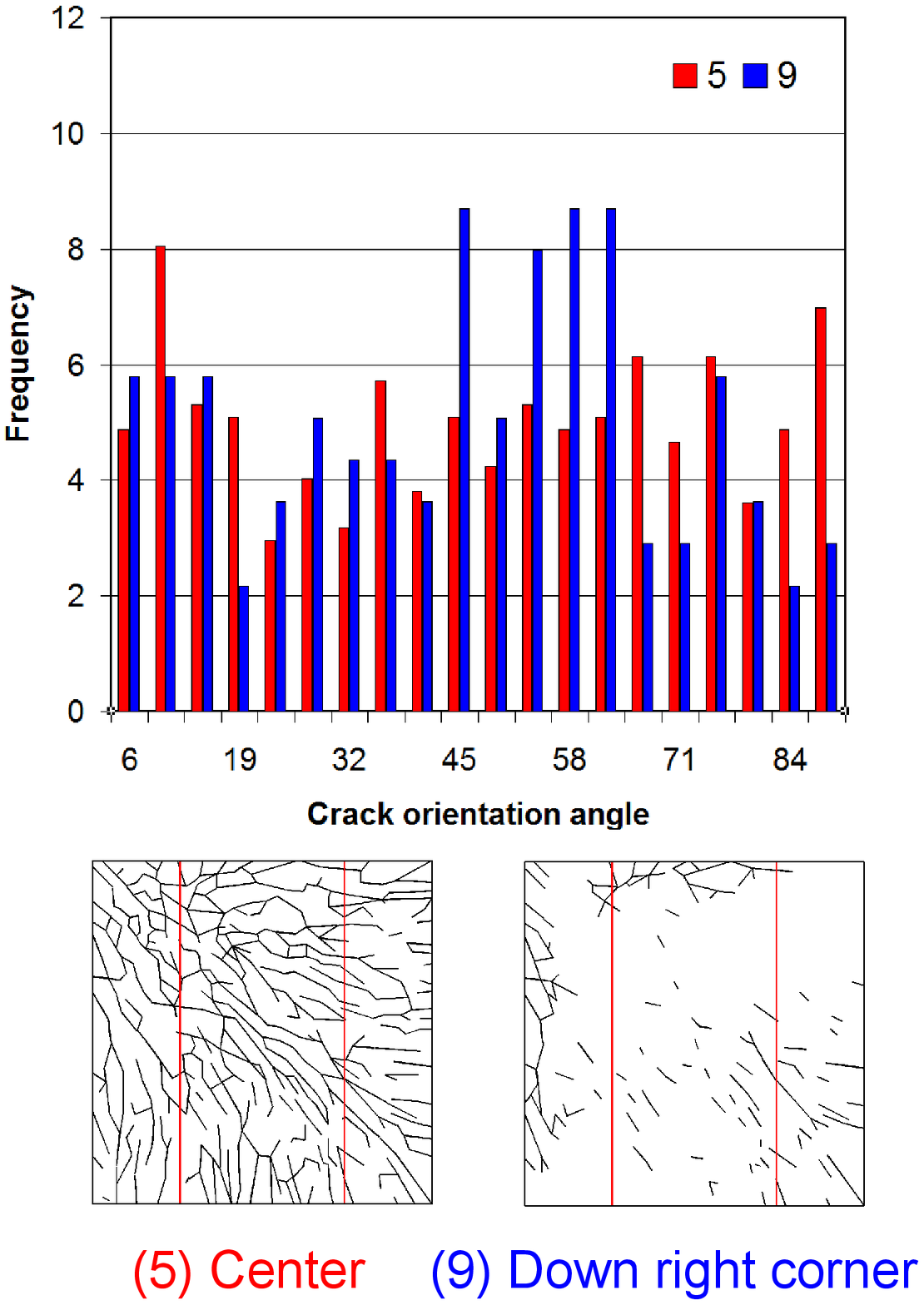}\label{fig13a}}}\quad
\centering \mbox{\subfigure[Left vs. right down corner cells]{\includegraphics[width=.44\textwidth,angle=0]{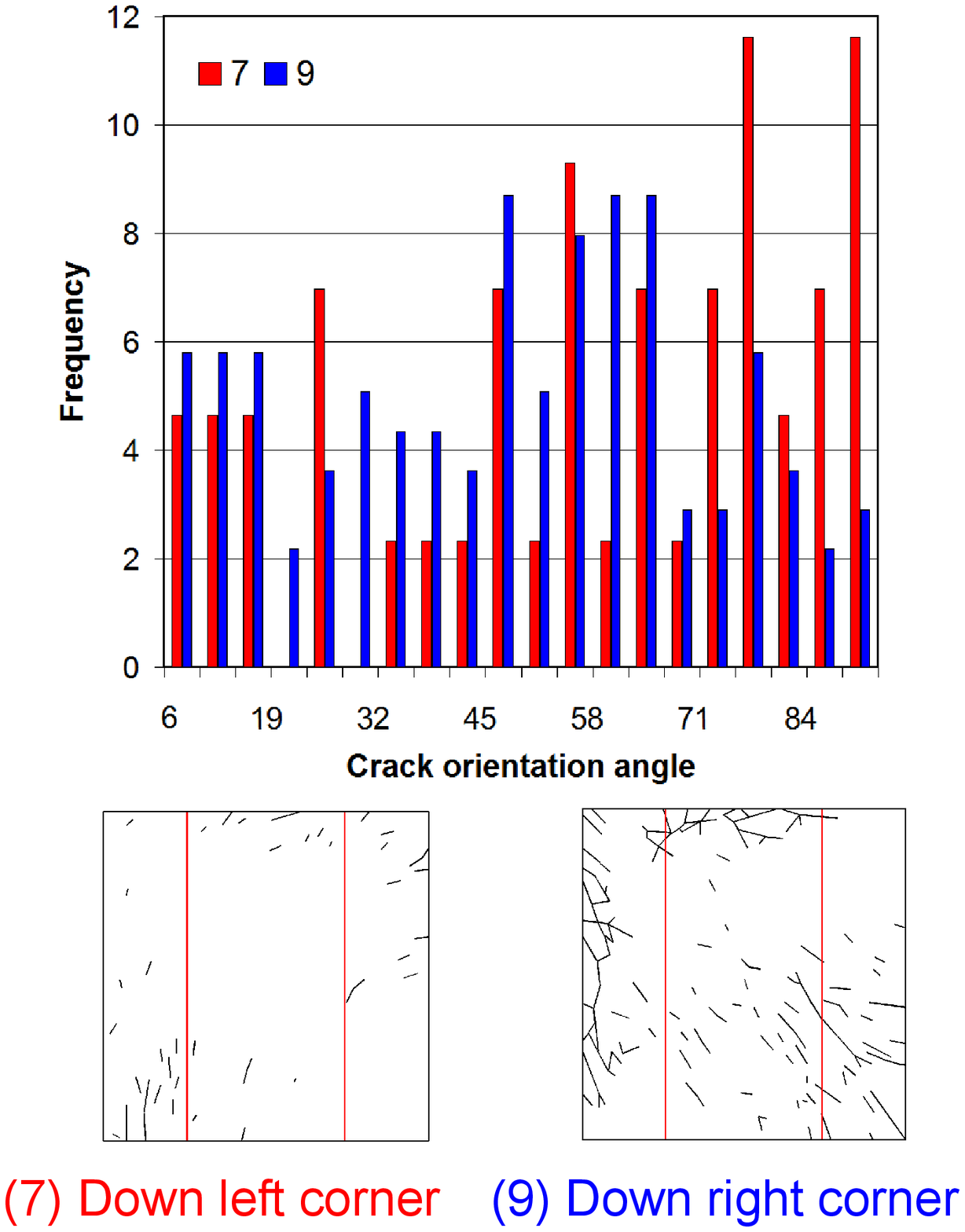}\label{fig13b}}}
\caption{Distribution of the orientation of microcracks:
comparison depending on the cell position. The angle $|\vartheta|=0^\circ$ denotes cracks perpendicular to the busbars (shown with red lines), whereas cracks with $|\vartheta|=90^\circ$ are parallel to the busbars.}\label{fig13}
\end{figure}

A comparison between the distribution of the orientations of microcracks inside the Si cells n. 5 (central cell) and 9 (lower right corner cell) is proposed in Fig.\ref{fig13a}. A quite uniform distribution is observed for the central cell, whereas a prevalence of microcracks with $|\vartheta|=45^{\circ}\div 60^{\circ}$ is observed for the corner cell. This is consistent with the experimental spatial and orientational distributions reported in \cite{KKEK}, where it was noticed that microcracks tend to align perpendicularly to the direction of the maximum principal tensile strain. In the plate corners, the maximum principal tensile strain is orientated at $|\vartheta|=45^{\circ}$ and it promotes the opening of the corresponding interface microcracks.

However, in addition to the influence of the direction of the macroscopic strain field, the orientation of Si grain boundaries might also have a role on the intergranular crack distribution. This is shown in Fig.\ref{fig13b}, where the orientational distribution of microcracks in cells n. 7 (lower left corner) and 9 (lower right corner) are compared. Since the macroscopic strain field is basically the same for all the corner cells, the difference is solely ascribed to the different grain boundary orientations. The microstructure used for all the cells shows in fact an evident elongation of the grains from the down right corner to the upper left corner. As a result of this, the Si cell n. 7 is less prone to microcracking in the $45^{\circ}$ direction than the cell n. 9.

\begin{figure}
\centering
\includegraphics[width=0.5\textwidth,angle=0]{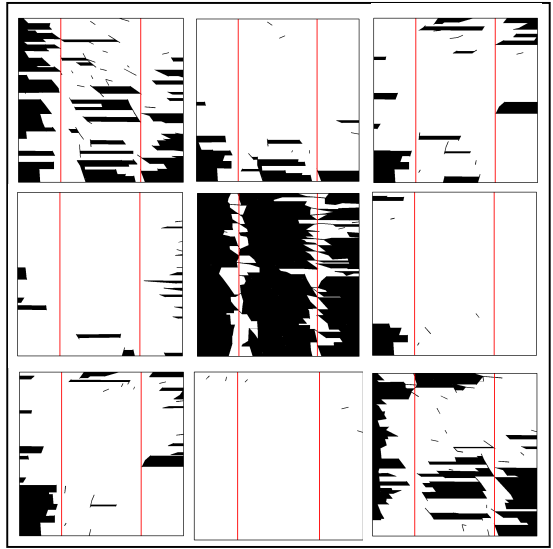}
\caption{Computed electrically inactive cell areas according to the microcrack pattern in Fig.\ref{fig12} (worst-case scenario in case of electrically insulated microcracks).} \label{fig14}
\end{figure}

The electrically inactive cell areas are also determined from the
crack pattern in Fig.\ref{fig12} and are shown in Fig.\ref{fig14}.
According to the criterion illustrated in Fig.\ref{fig7}, each cell
is subdivided in three distinct regions: the first on the left of
the first busbar, the second between the two busbars and the third
on the right of the second busbar. For the first region, the
inactive cell area is determined by the isolated cell area to the
left of the skyline given by the ensamble of microcracks closer to
the busbar. For the second region, the inactive cell area is
represented by the isolated area between two subvertical microcracks
with the same vertical coordinates. Finally, for the third region,
the inactive cell area is defined by all the cell areas to the right
of the skyline given by the ensemble of microcracks closer to the
busbar. This criterion should be considered as the worst case
scenario, since it is implicitly assumed that all the microcracks
are electrically insulated. Actually, partial conductivity in case
of closure effects due to temperature variations are likely to occur
and might be responsible for the oscillating electrical response of
defective cells, as experimentally observed in \cite{WSZB}.
According to the present assumptions, the damage variable for each
cell can be computed as the ratio between the black and the whole
cell areas. This can be done with a simple post-processing of the
cracked Si cell images in Matlab by computing the amount of black
and white pixels. The central cell is the most damaged with
$D=79\%$, whereas cells n. 1, 3, 7 and 9 have $D=32\%$, $12\%$,
$12\%$ and $30\%$, respectively. The less damaged cells are the n.
2, 4, 6 and 8, with $D=11\%$, $4\%$, $5\%$ and $\sim 0\%$,
respectively. The damage variable for the whole PV module is equal
to $D=79\%$.

\begin{figure}
\centering \mbox{\subfigure[Current vs.
voltage.]{\includegraphics[width=.4\textwidth,angle=0]{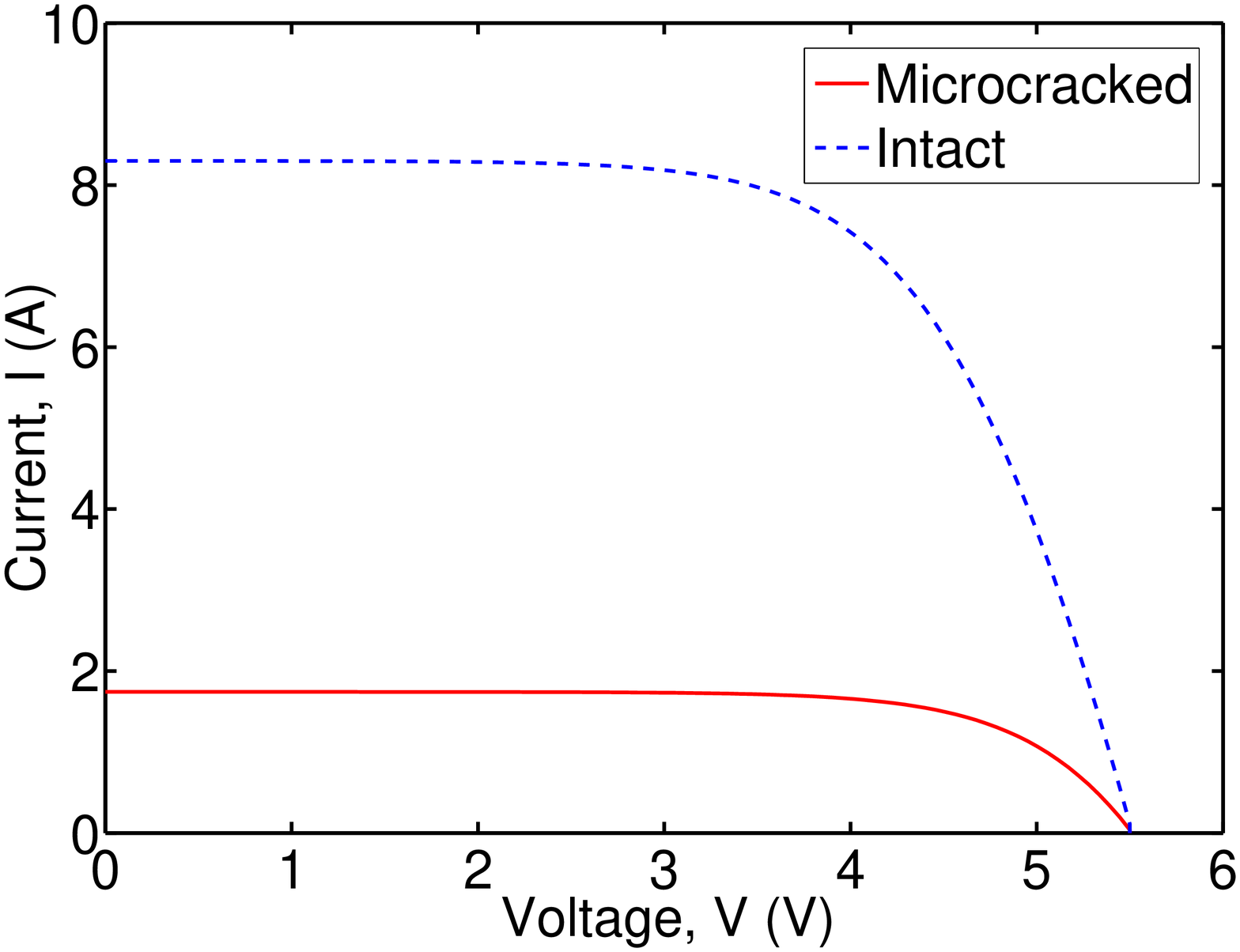}\label{fig15a}}}
\quad \centering \mbox{\subfigure[Power vs.
voltage.]{\includegraphics[width=.42\textwidth,angle=0]{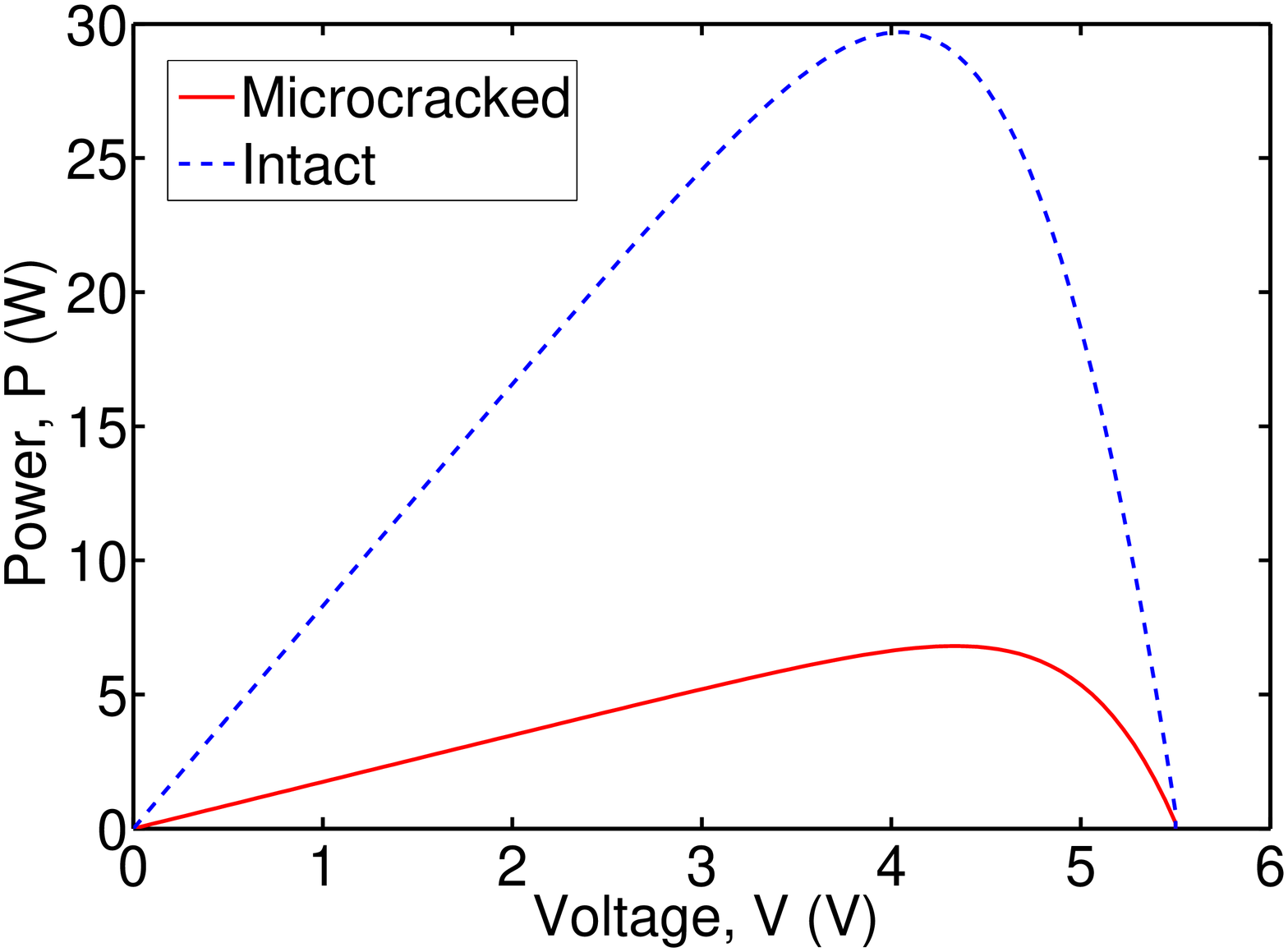}\label{fig15b}}}
\caption{Characteristic curves of the intact and micro-cracked PV
module: current vs. voltage (a) and power vs. voltage
(b).}\label{fig15}
\end{figure}

The electric response of the intact and microcracked PV modules
($a=2$, $I_{\text{ph}}^{D=0}=8.3$ A, $I_{\text{s}}^{D=0}=6\times
10^{-5}$ A, $R_{\text{s}}=0.007\,\Omega$,
$R_{\text{p}}=410\,\Omega$) is finally compared in Fig.\ref{fig15}.
For the computation we also consider an irradiance $I_r=1000$
W/m$^2$ and a working temperature $T=27^\circ$ (300 K). The effect
of microcracking is particularly evident and, in the present
worst-case scenario, the power loss is particularly high, with a
fill factor of the microcracked module reduced down to $15\%$ from
$65\%$ of the initial one.

\section{Conclusion}

In the present work, a multi-physics and multi-scale (multi-resolution) computational approach has been proposed for the study of the snow load-induced microcracking in polycrystalline Si solar cells and its effect on the electric response of PV modules. To the authors' best knowledge, this is the first computational approach that attempts at studying the coupling between the elastic and the electric fields. Moreover, it is the first computational method that explicitly considers the polycrystalline grain microstructure of Si using a multi-resolution approach which permits to study the structural response of a PV module without neglecting the role of the grain boundaries as a source of microcracking. The numerical application shows that the proposed approach can be applied to realistic case studies.

Future research in this field will regard both the quantitative
comparison with experimental measurements \cite{14}, parameter
identification, and the development of additional computational
features. In particular, the possibility of transgranular cracking,
not modelled in the present study, will be analyzed. The coupling
between the elastic and the thermal fields will also be put forward,
in order to address the important issue of durability of PV modules
exposed to cyclic temperature variations. The assumption of
perfectly insulating microcracks will also be checked with ad hoc
experiments and new constitutive micromechanical models for
partially conducting microcracks will be implemented. The
computationally efficient layer-wise mixed theories for laminated
plates proposed in \cite{carrera} will also be considered in
addition to 3D FE simulations.

\section*{Acknowledgements}
\noindent The research leading to these results has received funding
from the European Research Council under the European Union's
Seventh Framework Programme (FP/2007-2013) / ERC Grant Agreement n.
306622 (ERC Starting Grant "Multi-field and multi-scale
Computational Approach to Design and Durability of PhotoVoltaic
Modules" - CA2PVM). The support of the Italian Ministry of
Education, University and Research to the project FIRB 2010 Future
in Research "Structural mechanics models for renewable energy
applications" (RBFR107AKG) is also gratefully acknowledged. MP
thanks Dr. M. Köntges from the Institute of Solar Energy Research in
Hamelin, Germany, for the useful discussion about the electrical
model.

\bibliographystyle{unsrt}
%\bibliography{references}

%%%%%%%%%%%%%%%%%%%%%%%%%%%%%%%%%%%%%%%%%%%%%%%%%%%%%%%%%%%%%%%%%%%%%%

\end{document}